\documentclass[aps,pra,showpacs,twocolumn,preprintnumbers,byrevtex]{revtex4}
\usepackage{graphicx}

\begin{document}

\title{Fermat's principle of least time in the presence of
uniformly \\ moving boundaries and media}

\author{Aleksandar Gjurchinovski}

\email{agjurcin@iunona.pmf.ukim.edu.mk}

\author{Aleksandar Skeparovski}

\email{skepalek@iunona.pmf.ukim.edu.mk}

\affiliation{Department of Physics, Faculty of Natural Sciences 
and Mathematics, Sts.\ Cyril and Methodius University,
P.\ O.\ Box 162, 1000 Skopje, Macedonia}

\preprint{Eur. J. Phys. {\bf 28} (2007) 933--951}

\begin{abstract}

The refraction of a light ray by a homogeneous, isotropic 
and non-dispersive transparent material half-space in uniform 
rectilinear motion is investigated theoretically. The approach is 
an amalgamation of the original Fermat's principle and the fact 
that an isotropic optical medium at rest becomes optically 
anisotropic in a frame where the medium is moving at a constant 
velocity. Two cases of motion are considered: a) the material 
half-space is moving parallel to the interface; b) the material 
half-space is moving perpendicular to the interface. In each 
case, a detailed analysis of the obtained refraction 
formula is provided, and in the latter case, an intriguing 
backward refraction of light is noticed and thoroughly discussed. 
The results confirm the validity of Fermat's principle when the
optical media and the boundaries between them are moving at
relativistic speeds.

\end{abstract}


\date{December 06, 2007}

\maketitle

\section{Introduction}

The variational approach in geometrical optics was established
in 1657 when the famous French mathematician Pierre de Fermat  
provided a mathematical proof that the straight line is not the 
fastest way for light to traverse between two optical media \cite{dugas,hecht,born}. 
This simple and new way of looking at the propagation of light was 
already empirically noticed by Hero of Alexandria fifteen 
centuries before Fermat by observing that the path of the light between 
two points upon reflection from a plane mirror is the shortest 
possible path. Fermat's insightful and intriguing idea that ``the 
Nature always acts in the shortest ways'' has influenced generations 
of scientists, like Maupertius, Euler, Lagrange, Hamilton and 
Feynman, to successively translate classical and quantum physics 
into the language of variational calculus \cite{dugas,feynman1,feynman2,feynman3}.
Today, the practical value of Fermat's principle is immense, ranging 
from applications in geophysics, oceanography and material physics on 
one side, to general relativity and astrophysics on the 
other \cite{meshbey,nolet,berryman,shashidhar,cho,leonhardt1,leonhardt2,giannoni,
faraoni,kovner,leonhardt3,leonhardt4,leonhardt5,leonhardt6}.

The modern formulation of Fermat's principle states that a light 
in going between two fixed points traverses an optical path 
length that is stationary with respect to the variations of that 
path. In most cases, this formulation reduces to the original one that 
the actual path between the two points is the path which requires the 
shortest time \cite{hecht2}.
An accurate description of the process beyond Fermat's principle
is provided by the modern quantum electrodynamics, where the light is
considered to be made out of photons. Accordingly, the photon takes 
all the available paths in going between two fixed points, each path
at a certain probability, and the principle of least time follows as 
an approximation of this argument \cite{feynman4,feynman5,taylor}.

Fermat's principle has been initially established for optical systems 
whose properties are time-independent, and, as such, has entered the 
mainstream of the standard undergraduate physics curriculum through 
the derivations of the well-known laws of reflection and refraction of light 
at a smooth stationary interface between two optical materials at 
rest \cite{feynman4}. 
Recently, Fermat's principle was proven valid for optical systems with 
time-dependent parameters, incorporating the situations of light propagation
in uniformly moving optical media \cite{voronovich,godin1,godin2}. It was also 
shown that Einstein's law of reflection of light from a plane 
mirror in uniform rectilinear motion follows from Fermat's principle 
and the constant speed of light postulate, providing an indirect proof 
of the validity of Fermat's principle when the boundaries upon 
which the light is reflected are moving at relativistic speeds
\cite{gjurchinovski1}. Apart from the latter example, it seems that 
the problems of light propagation involving uniformly moving 
boundaries and media have never been attacked directly by 
using Fermat's principle of least time. 

Motivated by the latest extensions to non-stationary cases, in this 
paper we will use Fermat's principle to derive the formulas for 
the law of refraction of a light ray in two specific situations when the boundaries 
and the media are moving at constant velocities. We will limit our 
analysis to optical media that are homogeneous, isotropic and nonconducting 
in their rest frames of reference. Also, we consider the media to be 
non-dispersive at rest, in the sense that the correcting terms due 
to dispersion are so small that can be neglected. 

We begin our discussion in Sec. II by showing that an optical medium, 
isotropic in its rest frame, exhibits an optical anisotropy with respect 
to the frame where the medium is in uniform rectilinear motion. We derive 
the expression for the speed of the photon in the moving medium
and show that it depends on the angle between the light ray and the 
velocity of the medium. We describe this relativistically induced optical 
anisotropy in a more convenient way, by introducing an 
effective refractive index of the moving medium, defined as a ratio 
between the speed of the photon in vacuum and the speed of the photon 
in the moving medium. The notion that the photon will ``see'' the moving 
medium as a stationary medium having an effective refractive index 
different from its refractive index at rest, is incorporated into 
Fermat's principle in Sec. III and IV to investigate the refraction 
of the photon incident from a vacuum half-space upon a uniformly 
moving material half-space. In Sec. III we investigate the situation when the 
material half-space is moving parallel to the interface, and in Sec. IV we take 
the material half-space to move perpendicularly to the interface. In both cases, 
we consider the plane of incidence to be normal to the vacuum-material 
interface and parallel to the velocity of the medium.

A summary of the results and a comparison with other similar treatments
are given in Sec. V, where the possibility to extend the method to
more complicated configurations is also suggested.
Alternative derivation of the refraction law when the material half-space 
is moving perpendicular to the interface is presented in Appendix A.
Finally, in Appendix B, the conditions for the appearance of the backward 
refraction are derived analytically.

\section{An optical anisotropy of a transparent medium induced 
by its uniform motion}

If a flash of light is emitted from a fixed point source in an empty 
space (a vacuum), the outgoing photons will move radially on a spherical 
wavefront expanding at a speed of light $c$ \cite{sastry}. 
With respect to an observer from a frame that moves in a 
straight line at a constant velocity, the wavefront of the pulse 
will remain sphere, expanding at the same constant speed $c$ in all 
directions \cite{sastry2}. The result is a corollary of the second postulate of special relativity that 
the speed of light in a vacuum is the same in all inertial frames of reference 
regardless of the motion of the light source \cite{invariance}.

The situation is changed if the expansion of the pulse is taking
place in a material medium \cite{gjurchinovski2,schey}. 
If we consider a homogeneous, isotropic and non-dispersive
transparent medium at rest with respect to $S'$-frame, and
if we treat the problem two-dimensionally, the
equation that describes the evolution of the pulse is: 
\begin{equation}
x'^2+y'^2=(c/n)^2t'^2,
\label{eq:2.1}
\end{equation}
where $n$ is the refractive index of the material at rest.
This is an equation of an expanding circle whose radius at time
$t'$ is $(c/n)t'$. 
By using the Lorentz transformation:
\begin{equation}
x'=\gamma(x-ut),\
y'=y,\
t'=\gamma(t-ux/c^2),
\label{eq:2.2}
\end{equation}
with $\gamma=(1-u^2/c^2)^{-1/2}$, we can make a transition 
to $S$-frame where the medium is moving at a constant speed $u$ 
in the positive direction of the $x$-axis.
In this way, Eq. (\ref{eq:2.1}) is transformed into:
\begin{equation}
{(x-\eta t)^2\over (at)^2}+{y^2\over (bt)^2}=1,
\label{eq:2.3}
\end{equation}
where by $\eta$, $a$ and $b$ we denoted:
\begin{eqnarray}
\eta&=&u\left({1-1/n^2\over 1-u^2/n^2c^2}\right),
\label{eq:2.4}\\
a&=&(c/n)\left({1-u^2/c^2\over 1-u^2/n^2c^2}\right),
\label{eq:2.5}\\
b&=&(c/n)\left(1-u^2/c^2 \over 1-u^2/n^2c^2\right)^{1/2}.
\label{eq:2.6}
\end{eqnarray}
According to Eq. (\ref{eq:2.3}), the wavefront of the pulse
(the ray surface) with respect to an observer in $S$-frame is an 
expanding ellipse ``dragged'' by the moving medium (see Fig. 1).
\begin{figure}
\includegraphics[width=.45\textwidth,height=!]{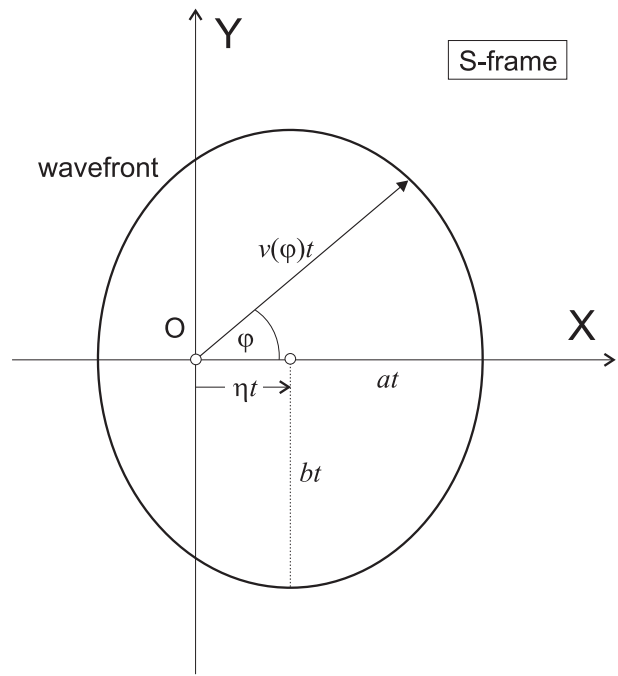}
\caption{An illustration of the light drag in the S-frame where the
medium is moving at a constant speed $u$ $(<c/n)$ in the direction of
the positive $x$-axis. The motion of the medium is a cause for 
the wavefront of the pulse originating from $O$ to be a dragged ellipse 
possessing an axial symmetry with respect to the $x$-axis. Here, the
photon emitted from $O$ at an angle $\varphi$ will travel at a speed 
$v(\varphi)$.}
\end{figure} 
At time $t$ from the emission of the pulse, the product $\eta t$ 
denotes the dragging parameter, that is, the distance from the 
origin $O$ of the pulse (the origin of the $xy$-coordinate system) 
to the center of the ellipse, and the products $at$ and $bt$ are 
the semi-minor and semi-major axes of the ellipse, respectively.
Consequently, the parameters $a$ and $b$ are the speeds of expansion 
of the semi-minor and semi-major axes of the ellipse, respectively, 
and $\eta$ is the speed of the center of the ellipse along the 
$x$-axis. 
When $\eta<a$, the ellipse will be incompletely ``dragged'', and the 
origin of the pulse will remain inside the ellipse at all
times. 
This is the case of a subluminal motion of the medium, when 
the speed of the medium $u$ is smaller than $c/n$. 
On the other hand, when $\eta\geq a$, the ``dragging'' is 
overwhelming, and the ellipse no longer encloses the origin. 
In this case the medium is moving at ``superluminal'' speeds ($c/n<u<c$),
causing an existence of a Mach cone, outside of which no 
light signal can ever reach, and inside of which light is doubled up,
with two pulses of light in each direction \cite{sastry,gjurchinovski2,schey}.

We may translate Eq. (\ref{eq:2.3}) into the polar coordinates by 
making the substitutions $x=vt\cos\varphi$ and $y=vt\sin\varphi$:
\begin{eqnarray}
\left( 1-{u^2\over c^2}\sin^2\varphi-{u^2\over n^2 c^2}
\cos^2\varphi\right)v^2 && \nonumber \\
-2u\cos\varphi\left( 1-{1\over n^2}\right)v+u^2-{c^2\over n^2}&=&0.
\label{eq:2.7}
\end{eqnarray}
Here, $v$ is the speed of the photon along a light ray 
directed at an angle $\varphi$ from the velocity of 
the medium (see Fig. 1).
Equation (\ref{eq:2.7}) is a quadratic in $v$, and it 
can be solved for $v$ in terms of the angle $\varphi$, the speed of 
the medium $u$, and the refractive index $n$ of the medium at rest.
By rejecting the second solution of Eq. (\ref{eq:2.7}) from the 
requirement that $v=c/n$ when $u=0$, we obtain:
\begin{eqnarray}
v(\varphi)&=&\biggl\{u(1-1/n^2)\cos\varphi+\bigl[c^2/n^2-u^2(\sin^2
\varphi \nonumber \\
&+&\cos^2\varphi/n^2)\bigr]^{1/2}\bigl[1-u^2/c^2\bigr]^{1/2}\biggr\}\nonumber\\
&\times&\biggl[1-u^2/c^2(\sin^2\varphi+\cos^2\varphi/n^2)\biggr]^{-1}.
\label{eq:2.8}
\end{eqnarray}
From Eq. (\ref{eq:2.8}) we conclude that the ray velocity in 
the moving medium depends on the angle $\varphi$ between the direction 
of the ray and the velocity of the medium. 
In this sense, an optical medium that is optically isotropic in 
its rest frame will possess an optical anisotropy in the frame in 
which it is moving at a constant velocity. 
This induced optical anisotropy is of a purely relativistic
origin, and it is different from the usual anisotropy in the
crystals \cite{bolotovskii}.

The observer in the $S$-frame may track the 
path of a light ray in the moving medium by considering it to be a
``stationary'' medium that possesses a relativistically induced optical 
anisotropy characterized by an effective refractive index $n_{ef}(\varphi)$:
\begin{equation}
n_{ef}(\varphi)={c\over v(\varphi)}.
\label{eq:2.9}
\end{equation}
Recalling Eq. (\ref{eq:2.8}) for $v(\varphi)$, the expression for $n_{ef}$
in Eq. (\ref{eq:2.9}) can be re-written in terms of $u$, $n$ and $\varphi$:
\begin{eqnarray}
n_{ef}(\varphi)&=&\biggl\{(u/c)(1-n^2)\cos\varphi+n[1-u^2/c^2]^{1/2} \nonumber \\
&\times&\bigl[(u/c)^2(n^2-1)\cos^2\varphi+1-n^2u^2/c^2\bigr]^{1/2}\biggr\}\nonumber\\
&\times&\left(1-n^2u^2/c^2\right)^{-1}.
\label{eq:2.10}
\end{eqnarray}
When $n=1$, Eqs. (\ref{eq:2.8}) and (\ref{eq:2.10})
reduce to $n_{ef}=1$ and $v=c$ for all $\varphi$, which describes an expansion of
the pulse in a vacuum. Also, if $u=0$, then $n_{ef}=n$ and $v=c/n$ for
all $\varphi$, which is the case when the medium is stationary. 

If the photon moves parallel to the velocity of the medium, $n_{ef}\approx n\pm u/c(1-n^2)$ 
and $v\approx c/n\pm u(1-1/n^2)$ to the first order in $u/c$, where the sign 
is taken ``+'' or ``--'' depending on whether $\varphi=0^\circ$ or 
$\varphi=180^\circ$, respectively. This longitudinal light drag has been theoretically 
predicted by Fresnel (1818) on the basis of his ether theory, and 
experimentally confirmed by Fizeau (1851) in his celebrated running-water 
experiment \cite{gjurchinovski2}.

\section{Refraction from a material half-space moving parallel to 
the interface}

A photon traveling in vacuum is incident upon the surface of a 
semi-infinite optical material which occupies the positive region
of the $y$-axis (see Fig. 2).
\begin{figure}
\includegraphics[width=.45\textwidth,height=!]{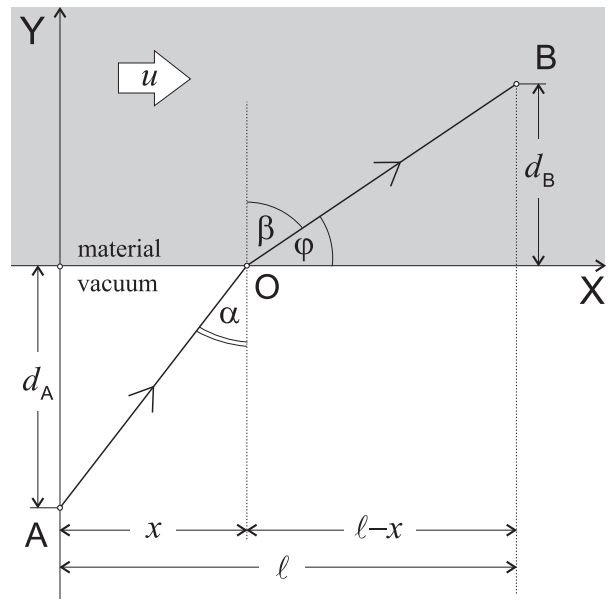}
\caption{The least-time derivation of the refraction formula when the material
half-space is moving at a constant speed $u$ in the direction of the
positive $x$-axis.}
\end{figure} 
The material half-space is moving at a constant speed $u$ in the 
positive direction of the $x$-axis, parallel to the vacuum-material interface.
The material in its rest frame has a refractive index $n$, and is assumed 
to be homogeneous, isotropic and non-dispersive. 
We will use Fermat's principle to derive the formula
for the law of refraction of the photon from the interface between the 
vacuum and the moving medium.

Let $A$ be the space-point in the vacuum half-space, belonging to the 
incident ray of the photon, and $B$ is the space-point in the moving 
medium that lies on the refracted ray.
By $d_A$ and $d_B$ we denote the shortest distances between the interface 
and the points $A$ and $B$, respectively, and $\ell$ is the distance 
between the orthogonal projections of the points $A$ and $B$ on the 
interface. 
According to Fermat's principle, the true path between the two fixed points 
$A$ and $B$ taken by the photon is the one that is traversed in the 
least time. 
Of all the points along the interface, the photon will be refracted at 
the point $O$ whose $x$-position minimizes the total time 
$t_{AB}$ required for the photon to cover the path from $A$ to $B$:
\begin{equation}
t_{AB}=t_{AO}+t_{OB}.
\label{eq:3.1}
\end{equation}
Here,
\begin{equation}
t_{AO}={\left(x^2+d_A^2\right)^{1/2}\over c}
\label{eq:3.2}
\end{equation}
is the time for the photon to travel from $A$ to $O$ at a speed of 
light $c$, and 
\begin{equation}
t_{OB}={\left[(\ell-x)^2+d_B^2\right]^{1/2}\over v(\varphi)}
\label{eq:3.3}
\end{equation}
is the corresponding time from $O$ to $B$. 
By $v(\varphi)$ we denote the speed of the photon along the refracted ray 
at an angle $\varphi$ from the velocity of the medium.
If we take into account that $v(\varphi)=c/n_{ef}(\varphi)$, and use 
Eqs. (\ref{eq:3.2}) and (\ref{eq:3.3}), we may re-write the total transit 
time $t_{AB}$ in Eq. (\ref{eq:3.1}) as:
\begin{equation}
t_{AB}={\left(x^2+d_A^2\right)^{1/2}\over c}+{n_{ef}(\varphi)\over 
c}\left[(\ell-x)^2+d_B^2\right]^{1/2}.
\label{eq:3.4}
\end{equation}    
We substitute $n_{ef}$ from Eq. (\ref{eq:2.10}) into Eq. (\ref{eq:3.4}),
use the relation 
\begin{equation}
\cos\varphi={\ell-x \over \left[(\ell-x)^2+d_B^2\right]^{1/2}}
\label{eq:3.5}
\end{equation}
from Fig. 2, and simplify the result to obtain:
\begin{eqnarray}
t_{AB}&=&{\left(x^2+d_A^2\right)^{1/2}\over c}+{(u/c^2)(1-n^2)\over 
1-n^2u^2/c^2}(\ell-x) \nonumber\\
&+&{(n/c)\left(1-u^2/c^2\right)^{1/2}\over 1-n^2u^2/c^2} \Biggl\{
(u/c)^2(n^2-1)(\ell-x)^2\nonumber \\
&+&(1-n^2u^2/c^2)\left[(\ell-x)^2+d_B^2\right]\Biggr\}^{1/2}.
\label{eq:3.6}
\end{eqnarray}
The principle of least time requires that $dt_{AB}/dx=0$, and thus:
\begin{eqnarray}
0&=&{x\over \left(x^2+d_A^2\right)^{1/2}}-{(u/c)(1-n^2)\over 1-n^2u^2/c^2}-
{n(1-u^2/c^2)^{3/2}\over 1-n^2u^2/c^2} \nonumber \\
&\times&{(\ell-x)\over \left[(\ell-x)^2+d_B^2\right]^{1/2}} 
\Biggl\{{(u/c)^2(n^2-1)(\ell-x)^2 \over \left[(\ell-x)^2+d_B^2\right]}\nonumber\\
&+&1-n^2u^2/c^2\Biggr\}^{-1/2}.
\label{eq:3.7}
\end{eqnarray} 
From Fig. 2 we notice the expressions for the angle of incidence $\alpha$ 
and the angle of refraction $\beta$ of the photon:
\begin{eqnarray}
\sin\alpha&=&{x\over \left(x^2+d_A^2\right)^{1/2}},
\label{eq:3.8}\\
\sin\beta&=&{\ell-x\over[(\ell-x)^2+d_B^2]^{1/2}},
\label{eq:3.9}
\end{eqnarray}
which we use into Eq. (\ref{eq:3.7}) to get:
\begin{eqnarray}
0&=&\sin\alpha-{(u/c)(1-n^2)\over 1-n^2u^2/c^2}-
{n(1-u^2/c^2)^{3/2}\over 1-n^2u^2/c^2}\nonumber \\
&\times&{\sin\beta\over \left[(u/c)^2(n^2-1)\sin^2\beta+1-n^2u^2/c^2\right]^{1/2}}.
\label{eq:3.10}
\end{eqnarray}
If we employ the identity $\sin\beta=\tan\beta(1+\tan^2\beta)^{-1/2}$ into 
Eq. (\ref{eq:3.10}) and do some algebra, we obtain a quadratic 
equation in $\tan\beta$:
\begin{eqnarray}
\left[n^2\left(1-{u\over c}
\sin\alpha\right)^2-\left(\sin\alpha-{u\over c}\right)^2\right] 
\tan^2\beta&-& {1\over 1-u^2/c^2} \nonumber \\
\times\left[\left(1-{n^2u^2\over c^2}\right)\sin\alpha+{u\over c}(n^2-1)
\right]^2&=&0.
\label{eq:3.11}
\end{eqnarray}
Equation (\ref{eq:3.11}) has two solutions in $\tan\beta$, and the 
solution: 
\begin{eqnarray}
\tan\beta&=&\bigl[(n^2-1)(u/c)+(1-n^2u^2/c^2)\sin\alpha\bigr] \nonumber \\
&\times&\biggl\{n^2\bigl[1-(u/c)\sin\alpha \bigr]^2-(\sin\alpha-u/c)^2\biggr\}^{-1/2} \nonumber\\
&\times&\left(1-u^2/c^2\right)^{-1/2}
\label{eq:3.12}
\end{eqnarray}
is the only one that is physically correct, which we choose from 
the requirement that when $u=0$ it should reduce to the usual Snell's 
law of refraction.

Equation (\ref{eq:3.12}) is the formula for the law of refraction of 
the photon from the vacuum-material interface when the optical material
is moving uniformly parallel to its surface.
It gives the angle of refraction $\beta$ of the photon as a 
function of the incident angle $\alpha$, the speed $u$ of the medium, 
and the refractive index $n$ of the medium in its rest frame. 
[Two alternative derivations leading to the same refraction formula 
are given in Ref. \cite{gjurchinovski2}.]

The functional dependence of the angle of refraction $\beta$ on the 
incident angle $\alpha$ is plotted in Fig. 3 for $n=1.5$ and for 
several different values of the speed $u$ of the medium.
\begin{figure}
\includegraphics[width=.45\textwidth,height=!]{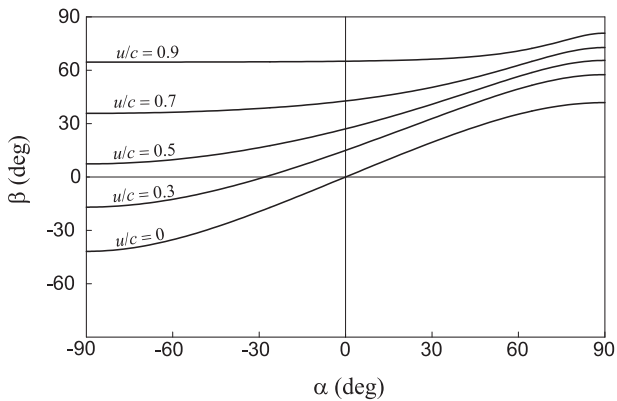}
\caption{The angle of refraction $\beta$ versus the incident angle $\alpha$
for $n=1.5$ and for different speeds $u$ of the material half-space, when the 
material half-space is moving parallel to the interface.}
\end{figure} 
We immediately notice that the law of refraction in Eq. (\ref{eq:3.12}) 
differs considerably from the usual Snell's law in the case of a stationary
medium.
First of all, when $u/c>0$, the angle of refraction $\beta$ attain non-zero values
even when the incident angle $\alpha$ is zero. 
In other words, the photon that enters the moving medium perpendicularly
to its surface will be dragged by the medium in the moving direction.
This is the transverse Fresnel-Fizeau light drag, theoretically
predicted by Fresnel in 1818 and experimentally detected by Jones in the
early 1970s with an accuracy up to the first order 
of $u/c$ \cite{jones1,jones2,gjurchinovski2,dispersive}. 

The influence of the dragging effect on the refraction is also
visible for non-zero incident angles, especially when they take 
negative values \cite{negative}. 
In the latter case, if $u/c>0$, starting from $\alpha=0^\circ$ down to a certain 
negative incident angle, the angle of refraction $\beta$ is positive 
(see Fig. 3), which means that both the incident and the refracted ray of the photon will 
lie on the same side with respect to the interface normal.  
Moreover, when the speed $u$ of the medium becomes greater than $c/n^2$, the
angle of refraction will be positive for any value of the incident angle 
$\alpha$.

Equation (\ref{eq:3.12}) can also be used to calculate the angle of refraction
of the photon when the medium moves in the opposite direction to the one in Fig. 2, 
that is, in the negative direction of the $x$-axis. 
However, the situation when the photon is incident at a positive angle $\alpha$
upon the medium which moves at a speed $-u$ is physically identical to the
situation which corresponds to a negative incident angle $-\alpha$ and a speed $u$.

At the end, we emphasize that the absolute value of the angle of 
refraction $\beta$ in Eq. (\ref{eq:3.12}) cannot become equal or greater than 
$90^\circ$, no matter what the values of $\alpha$, $n$ and $u$ are. 
In other words, it is not possible to observe the situation in which the photon
would slide along the surface of the medium, or to be reflected back to the 
vacuum half-space. 
To prove this property, we will try to find such a value of $\alpha$
which will turn the denominator in the right-hand side of Eq. (\ref{eq:3.12})
into zero. 
This is the condition for the angle $\beta$ to equal $\pm 90^\circ$. 
Hence, $\alpha$ should be such that:
\begin{equation}
n^2\left(1-{u\over c}\sin\alpha\right)^2-\left(\sin\alpha-{u\over c}\right)^2=0,
\label{eq:3.13}
\end{equation}
which is a quadratic equation in $\sin\alpha$, and has two solutions:
\begin{equation}
(\sin\alpha)_{1,2}={n\pm u/c \over nu/c\pm 1}.
\label{eq:3.14}
\end{equation}
Having in mind that $n>1$ and $u<c$, we have 
$\vert (n\pm u/c)(nu/c\pm 1)^{-1}\vert>1$, which means that there are no
real values for the angle $\alpha$ to satisfy Eq. (\ref{eq:3.13}).

\section{Refraction from a material half-space moving perpendicular to 
the interface}

In the following, we consider the case in which the material 
half-space in Sec. III is moving at a constant speed $u$ in the 
positive direction of the $y$-axis. 
The geometry of the problem is shown in Fig. 4.
\begin{figure}
\includegraphics[width=.45\textwidth,height=!]{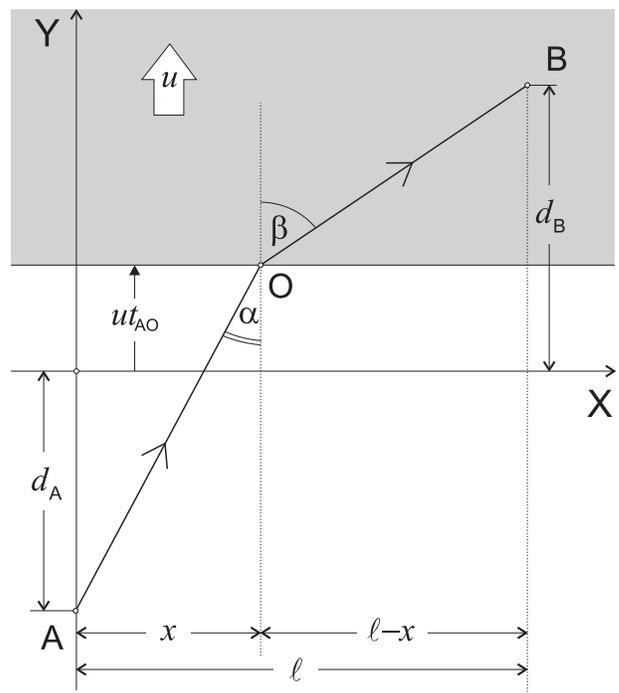}
\caption{The least-time derivation of the refraction formula when the material
half-space is moving at a constant speed $u$ in the positive direction of the
$y$-axis.}
\end{figure} 
Observe that unlike the case in Sec. III, the vacuum-material interface 
also is in uniform motion, having the same constant speed $u$ as the
medium in the positive $y$-direction.

A photon is emitted from point $A$ in the vacuum half-space. 
At this instant, the vacuum-material interface coincides with the 
$x$-axis, and $d_A$ is the shortest distance between the interface 
and the point $A$. 
After being refracted from the moving interface, the photon reaches 
the fixed point $B$ in the material half-space.
We denote by $d_B$ the shortest distance between the interface 
and the point $B$ at time when the photon was emitted from $A$, and 
by $\ell$ the distance between the orthogonal projections of the 
points $A$ and $B$ on the interface.

Due to the motion of the vacuum-material interface, the hypothetical 
points from which the photon may be refracted at the interface will not 
lie on the same horizontal.
According to Fermat's principle, the photon will be refracted from a 
point $O$ that makes $AOB$ the path of the shortest transit time (see Fig. 4).
We denote by $t_{AO}$ the time required for the photon to go from $A$ to
$O$, and by $t_{OB}$ the time from $O$ to $B$. 
Observe that when the photon reaches point $O$, the interface has already 
crossed the distance $ut_{AO}$ in the positive $y$-direction from its 
position $y=0$ when the photon was emitted from $A$. 

The total time $t_{AB}$ for the photon to travel the path $AOB$ is:
\begin{equation}
t_{AB}=t_{AO}+t_{OB}.
\label{eq:4.1}
\end{equation}
From Fig. 4 we have:
\begin{eqnarray}
(ct_{AO})^2&=&(d_A+ut_{AO})^2+x^2, 
\label{eq:4.2} \\
\left[v(\beta)t_{OB}\right]^2&=&(d_B-ut_{AO})^2+(\ell-x)^2,
\label{eq:4.3} 
\end{eqnarray}
where $v(\beta)$ is the speed of the photon in the moving medium in 
the direction determined by the angle $\beta$. 
The speed $v(\beta)$ of the photon follows from Eq. 
(\ref{eq:2.8}) if we simply substitute $\beta$ for $\varphi$.
In order to simplify the derivation of the refraction formula, we 
re-write Eqs. (\ref{eq:4.2}) and (\ref{eq:4.3}) as:
\begin{eqnarray}
t_{AO}&=&{1\over c}\left[(d_A+ut_{AO})^2+x^2\right]^{1/2}, 
\label{eq:4.4} \\
t_{OB}&=&{n_{ef}(\beta)\over c}\left[(d_B-ut_{AO})^2+(\ell-x)^2\right]^{1/2},
\label{eq:4.5} 
\end{eqnarray}
with $n_{ef}(\beta)$ being the effective refractive index of the moving
half-space in this case: 
\begin{eqnarray}
n_{ef}(\beta)&=&\biggl\{(u/c)(1-n^2)\cos\beta+n[1-u^2/c^2]^{1/2} \nonumber \\
&\times&\bigl[(u/c)^2(n^2-1)\cos^2\beta+1-n^2u^2/c^2\bigr]^{1/2}\biggr\}\nonumber\\
&\times&\left(1-n^2u^2/c^2\right)^{-1}.
\label{eq:4.6}
\end{eqnarray}
Thus, the total transit time $t_{AB}$ in Eq. (\ref{eq:4.1}) can be re-written as:
\begin{eqnarray}
t_{AB}&=&{1\over c}\left[(d_A+ut_{AO})^2+x^2\right]^{1/2} \nonumber \\
&+&{n_{ef}(\beta)\over c}\left[(d_B-ut_{AO})^2+(\ell-x)^2\right]^{1/2}.
\label{eq:4.7} 
\end{eqnarray}
If we substitute Eq. (\ref{eq:4.6}) into Eq. (\ref{eq:4.7}), and take into
account the expression:
\begin{equation}
\cos\beta={d_B-ut_{AO}\over \left[(d_B-ut_{AO})^2+(\ell-x)^2\right]^{1/2}},
\label{eq:4.8}
\end{equation}
from Fig. 4, we obtain an expression for the transit time $t_{AB}$ of 
the photon as a function of the position $x$
of the point of refraction $O$ at the moving interface:
\begin{eqnarray}
t_{AB}&=&{\left[(d_A+ut_{AO})^2+x^2\right]^{1/2}\over c}+{u(1-n^2)(d_B-ut_{AO})\over c^2-n^2u^2} \nonumber \\
&+&{(n/c)\left(1-u^2/c^2\right)^{1/2}\over 1-n^2u^2/c^2}
\Biggl\{{u^2\over c^2}(n^2-1)(d_B-ut_{AO})^2 \nonumber \\
&+&\left(1-{n^2u^2\over c^2}
\right)\left[(d_B-ut_{AO})^2+(\ell-x)^2\right]\Biggr\}^{1/2}.
\label{eq:4.9}
\end{eqnarray}
We now invoke the principle of least time that the value of $x$
should be such that the transit time $t_{AB}$ of the photon is minimum. We take
the derivative of $t_{AB}$ in Eq. (\ref{eq:4.9}) with respect to $x$ and set the
result to zero to obtain:
\begin{eqnarray}
0&=&u\cos\alpha\left({dt_{AO}\over dx}\right)+\sin\alpha+{(u^2/c)(n^2-1)\over 1-n^2u^2/c^2}\nonumber \\
&\times&\left({dt_{AO}\over dx}\right)-{n\left(1-u^2/c^2\right)^{1/2}\over 1-n^2u^2/c^2} 
\Biggl[u\left(1-{u^2\over c^2}\right) \nonumber \\
&\times&\left({dt_{AO}\over dx}\right)\cos\beta+\left(1-{n^2u^2\over c^2}\right)\sin\beta\Biggr]\nonumber \\
&\times&\Biggl[{u^2\over c^2}(n^2-1)\cos^2\beta+1-{n^2u^2\over c^2}\Biggr]^{-1/2},
\label{eq:4.10}
\end{eqnarray}
where we employed the relations:
\begin{eqnarray}
\sin\alpha&=&{x \over \bigl[x^2+(d_A+ut_{AO})^2\bigr]^{1/2}}, 
\label{eq:4.11}\\
\cos\alpha&=&{d_A+ut_{AO} \over \bigl[x^2+(d_A+ut_{AO})^2\bigr]^{1/2}},
\label{eq:4.12}\\
\sin\beta&=&{\ell-x \over \bigl[(\ell-x)^2+(d_B-ut_{AO})^2\bigr]^{1/2}},
\label{eq:4.13}
\end{eqnarray}
from Fig. 4, together with Eq. (\ref{eq:4.8}) and the fact that 
$t_{AO}=t_{AO}(x)$. 
To find $dt_{AO}/dx$, we take the derivative of Eq. (\ref{eq:4.2}) with
respect to $x$ and obtain:
\begin{equation}
2c^2t_{AO}\left({dt_{AO}\over dx}\right)=2(d_A+ut_{AO})u
\left({dt_{AO}\over dx}\right)+2x,
\label{eq:4.14}
\end{equation}
from which we get:
\begin{equation}
\left({dt_{AO}\over dx}\right)={x/(ct_{AO}) \over c\left[1-(u/c)(d_A+ut_{AO})
/(ct_{AO})\right]}.
\label{eq:4.15}
\end{equation}
By recognizing from Fig. 4 that $\sin\alpha=x/(ct_{AO})$ and 
$\cos\alpha=(d_A+ut_{AO})/(ct_{AO})$, we recast Eq. (\ref{eq:4.15}) 
into the form:
\begin{equation}
\left({dt_{AO}\over dx}\right)={\sin\alpha \over c-u\cos\alpha}.
\label{eq:4.16}
\end{equation}
We substitute Eq. (\ref{eq:4.16}) into Eq. (\ref{eq:4.10}), apply
the trigonometric identities $\sin\beta=\tan\beta(1+\tan^2\beta)^{-1/2}$ 
and $\cos\beta=(1+\tan^2\beta)^{-1/2}$, and do some algebra to obtain
a quadratic equation in $\tan\beta$:
\begin{eqnarray}
\left\{{n^2(1-n^2u^2/c^2)\over 1-u^2/c^2}-{\sin^2\alpha \over 
\left[1-(u/c)\cos\alpha\right]^2}\right\}\tan^2\beta
\nonumber\\
+{2n^2(u/c)\sin\alpha \over 1-(u/c)\cos\alpha}\tan\beta 
-{(1-u^2/c^2)\sin^2\alpha\over \left[1-(u/c)\cos\alpha\right]^2}=0.
\label{eq:4.17}
\end{eqnarray}
Equation (\ref{eq:4.17}) has two solutions in $\tan\beta$, 
and its only physically correct solution is  
\begin{eqnarray}
\tan\beta=(1-u^2/c^2)\sin\alpha\times \biggl\{n^2(u/c)\left[1-(u/c)\cos\alpha\right] \nonumber \\
+\Bigl[n^2\left[1-(u/c)\cos\alpha\right]^2-(1-u^2/c^2)\sin^2\alpha\Bigr]^{1/2}\biggr\} ^{-1}, \nonumber \\
\label{eq:4.18}
\end{eqnarray}
which is the formula for the law of refraction of the photon in this case.
We reject the second solution of Eq. (\ref{eq:4.17}) as a physically incorrect
one, by using the same argument as in the deduction of Eq. (22) from Eq. (21). 
[A derivation of Eq. (\ref{eq:4.18}) via the relativistic velocity transformation 
formulas is provided in Appendix A.] 

The obtained refraction formula works both for positive and negative values of the 
incident angle $\alpha$, as well as for both moving directions of the medium.
The right-hand side of Eq. (\ref{eq:4.18}) is an odd function of $\alpha$, and hence
the change in the sign of $\alpha$ does not introduce physically new situation.
The path of the photon which is incident upon the surface at a
negative angle $-\alpha$ will be symmetric with respect to the surface normal, to the
path of the photon corresponding to a positive incident angle $\alpha$.
Therefore, in the following, we consider only positive incident angles 
($0^\circ\leq\alpha <90^\circ$).

When the medium moves in the positive direction of the $y$-axis, $u/c>0$ and
the right-hand side of Eq. (\ref{eq:4.18}) is always positive, which means that
the angle of refraction $\beta$ is between $0^\circ$ and $90^\circ$. 
The dependence of $\beta$ on the incident angle $\alpha$ for $n=1.5$ and for
various positive values of $u/c$ is shown in Fig. 5.
\begin{figure}
\includegraphics[width=.45\textwidth,height=!]{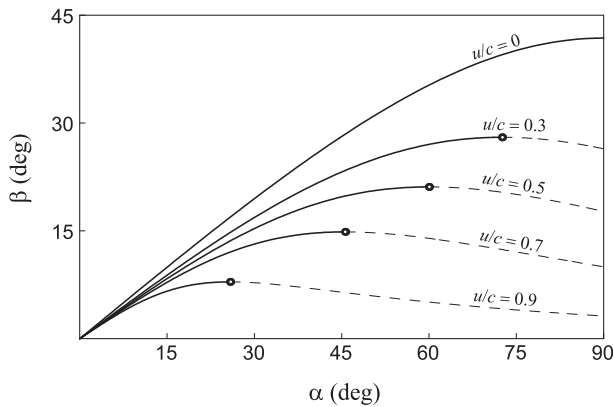}
\caption{The angle of refraction $\beta$ versus the incident angle $\alpha$
for $n=1.5$ and for different speeds $u$ of the material half-space 
moving in the positive direction of the $y$-axis. The portions of the
curves corresponding to $\alpha\geq\arccos(u/c)$ are drawn with dashed lines.
At incident angles corresponding to the dashed curves, the refraction does not 
occur.}
\end{figure} 
We see that as $u/c$ varies from $0$ to $1$, the curves corresponding to
different values of $u/c$ are regularly shifted toward the smaller values
of $\beta$, which is expectable if one takes into account the dragging effect of the
moving medium. We emphasize that the portions of the curves drawn with dashed lines
correspond to incident angles at which the photon will never reach the 
interface. Namely, for a given speed $u$ of the medium in the positive $y$-direction,
there exists an interval of values for the incident angle $\alpha$ ($\alpha_{\text{max}}
\leq\alpha<90^\circ$) for which the $y$-projection of the photon's velocity ($c\cos\alpha$)
is less than or equal to the velocity of the medium. The value of $\alpha_{\text{max}}$ 
follows from:
\begin{equation}
\cos\alpha_{\text{max}}=u/c,
\label{eq:4.19}
\end{equation}
and it can be shown that it is exactly the incident angle at which the curve $\beta=\beta(\alpha)$
attains the maximum value. Hence, for incident angles belonging to the interval 
$[\alpha_{\text{max}},90^\circ]$, the refraction becomes impossible. Translating to the
frame in which the medium is at rest, and thus taking into account the aberration 
phenomenon, we have a photon ``incident'' upon the interface at angles larger than $90^\circ$.

The situation changes dramatically when $u/c<0$. 
In this case, the denominator on the right-hand side of Eq. (\ref{eq:4.18})
can take zero or even negative values, which implies that the angle of
refraction $\beta$ can become equal or greater than $90^\circ$. 
In Fig. 6 we plotted $\beta$ as a function of $\alpha$ for $n=1.5$ and for
various negative values of $u/c$.
\begin{figure}
\includegraphics[width=.45\textwidth,height=!]{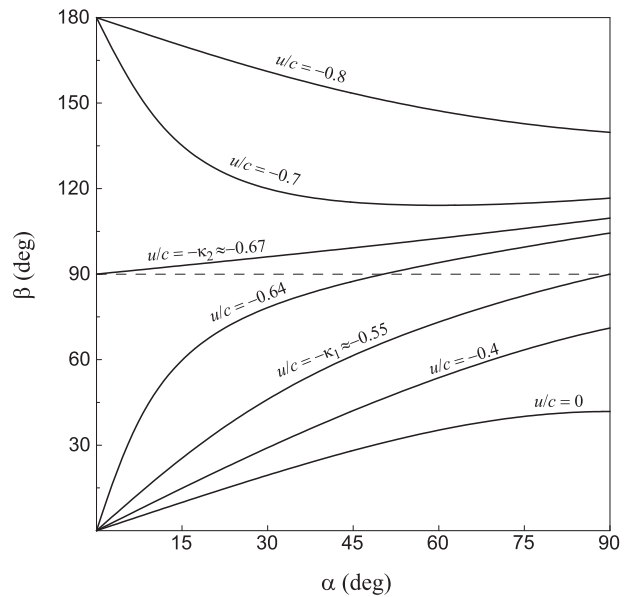}
\caption{The angle of refraction $\beta$ as a function on the incident angle $\alpha$
for $n=1.5$ when the material half-space is moving perpendicular to the interface, 
in the negative direction of the $y$-axis. When $-\kappa_1<u/c<0$ the refraction is
regular, and the angle of refraction $\beta$ of the photon is in the interval 
$[0^\circ,90^\circ]$. For $-\kappa_2<u/c<-\kappa_1$, the refraction is:
a) regular ($0\leq\beta<90^\circ$), if $0\leq\alpha<\alpha_c$; and b) backward ($90<\beta<180$), 
if $\alpha_c<\alpha<90^\circ$. If $u/c<-\kappa_2$, the refraction is always backward.} 
\end{figure} 
What we see in Fig. 6 can be summarized as follows.
When the medium moves in the negative direction of the $y$-axis up to some value
of $\vert u/c\vert$ denoted as $\kappa_1$, the refraction seems to be regular.
In this range of speeds ($-\kappa_1<u/c<0$), as the absolute value of the speed increases, the
curves are shifted toward higher values of $\beta$, and $\beta$ remains smaller
than $90^\circ$ for each $\alpha$. 
When $u/c<-\kappa_1$, a specific value for the incident angle $\alpha$ appears for 
which the angle of refraction of the photon becomes equal to $90^\circ$. 
We denote this critical incident angle by $\alpha_c$.
For $\alpha_c<\alpha<90^\circ$, the angle of refraction $\beta$ is greater than $90^\circ$.
By further increasing the absolute value of $u/c$, the critical angle $\alpha_c$
decreases, and for some value of $\vert u/c\vert$, denoted by $\kappa_2$, 
$\alpha_c$ becomes $0^\circ$.
For $u/c<-\kappa_2$, the angle of refraction $\beta$ is greater 
than $90^\circ$ for all values of $\alpha$.
The quantities $\kappa_1$ and $\kappa_2$ are functions on the refractive index $n$ of the medium 
in its rest frame, and as $n$ increases, the values of $\kappa_1$ and $\kappa_2$ are becoming 
closer to each other.
It can be shown (see Appendix B) that $\kappa_1=(1+n^2)^{-1/2}$ and $\kappa_2=1/n$ 
(in our case, for $n=1.5$, $\kappa_1\approx 0.55$ and $\kappa_2\approx 0.67$).
Evidently, the value $\kappa_2$ corresponds to the smallest superluminal speed of the medium.
\begin{figure*}
\includegraphics[width=.65\textwidth,angle=0,clip]{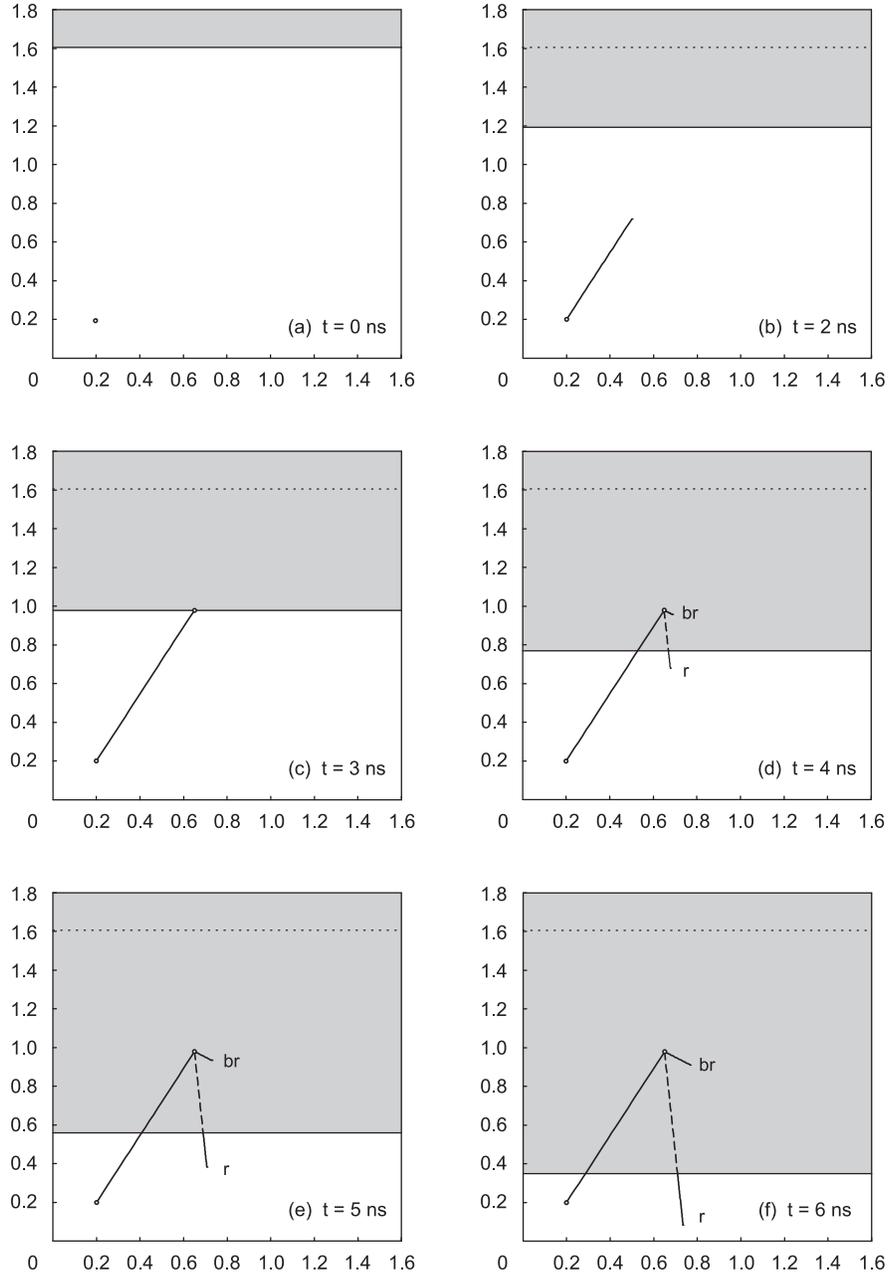}
\caption{A space-time simulation depicting the difference between the reflection and the 
backward refraction. The material half-space (the shaded region) is moving in the negative 
$y$-direction at a speed $0.7c$. The refractive index of the material half-space in its rest 
frame is $n=1.5$. The values of $x$ and $y$ coordinates are given in meters.  
(a) The situation at time $t=0\text{ ns}$. A photon, located at point $(0.2,0.2)$, is incident
upon the vacuum-material interface, making an angle of $30^\circ$ with respect to the interface normal
in the vacuum half-space. The interface is located at $y=1.6084478$; (b) The path of the photon
and the position of the interface at time $t=2\text{ ns}$;
(c) The situation at $t=3\text{ ns}$. The photon reaches the interface at point $(0.649689,0.97884)$;
(d) The situation at $t=4\text{ ns}$. The ray $br$ corresponds to the path of the backwardly
refracted photon, making an angle of $119.974^\circ$ with respect to the interface normal
in the material half-space. The speed of the backwardly refracted photon is $0.151c$, and its $y$ 
velocity component is $-0.07544c$. The ray $r$ is the path of the 
photon if it were reflected from the interface. The angle of reflection is $-5.41445^\circ$    
with respect to the interface normal in the vacuum half-space. The $y$ velocity component of the
reflected photon is $-0.99554c$; (e) The situation at $t=5\text{ ns}$; (f) The situation at 
$t=6\text{ ns}$.}
\end{figure*} 

The existence of angles of refraction greater than $90^\circ$ in the case $u/c<0$ implies 
that the photon, instead of moving in the same general direction as before the refraction, 
moves backward. At first sight, it might seem that the photon was reflected at the interface 
instead of being refracted, but this is not what had happened (see Fig. 7). Once the photon strikes 
the vacuum-material interface, it undergoes a refraction, enters the medium and continues 
to move into the medium at all times. The curious behavior of the refracted photon is due 
to the fact that the photon, while penetrating into the medium, is dragged by the overall 
motion of the medium. This dragging effect is a cause for the $y$-component of the 
photon's velocity in the moving medium to be negative, and hence, the refracted photon will move in the 
same general direction as the medium.

We emphasize that the path of the photon that undergoes a backward refraction does not
coincide with the path which it would trace if it were reflected from the interface (see Fig. 7).
The path of the backwardly refracted photon is fully determined by Eq. (\ref{eq:4.18}), 
and the path of the reflected photon obeys the Einstein's formula for reflection from 
a uniformly moving mirror (see Ref. \cite{einstein,gjurchinovski1,gjurchinovski3}).
In the language of modern quantum electrodynamics, whether the photon will be reflected or 
refracted at the inerface is a matter of probability. In this sense, if a beam
of photons is incident on the moving interface at an angle $\alpha$, and if the conditions
for the appearance of the backward refraction are fulfilled, some of the photons
will be reflected, and some backwardly refracted.
Therefore, in a more realistic view in which the reflection is considered also, two
different beams spreading backward will be identified.
In this way, we have a full consistency between the observations with respect to both
inertial frames, $S$ and $S'$. 
Namely, for the observer in $S'$-frame to which the medium is at rest, both processes
are developing in an ordinary way: the refracted photon moves forward, obeying the Snell's 
law of refraction, while the reflected one moves backward, according to the usual law of 
reflection: incident angle equals the reflected angle. 
Hence, the observer in $S'$ clearly distinguishes two rays, and consequently, two rays must
exist with respect to the observer in the $S$-frame: one reflected, and one backwardly refracted.

\section{Concluding remarks}

We have presented a novel approach to the problem of refraction 
of a light ray at an interface between two homogeneous, isotropic and non-dispersive
transparent optical materials in uniform rectilinear motion. 
Our method amalgamates the original Fermat's principle and the fact that an
isotropic optical material at rest becomes optically anisotropic if it is moving 
at a constant velocity. The derivation is in the framework of basic optics and 
special relativity, requiring some knowledge of calculus at an elementary level.
As such, it may be regarded as an instructive addendum to the typical introductory 
physics courses.

We have analyzed in details the refraction at a vacuum-material 
interface, when the material half-space is moving uniformly parallel or uniformly 
perpendicular to the interface. In both cases, we considered the plane of incidence to be 
normal to the interface and parallel to the velocity of the 
medium. The coincidence of the obtained refraction formulas with the ones obtained 
with the other methods confirms the validity of Fermat's principle in the presence
of uniformly moving boundaries and media.

In addition, when the material half-space moves perpendicular to the 
interface, in the direction which is opposite to that of the incident light, 
we observed a ``backward refraction'' of light. To our knowledge, this optical effect 
has not been noticed before in the literature. 

We emphasize that the refraction formulas obtained in this paper refer to 
refraction of a photon, and thus, describe the refraction of a light ray. 
We note that in a set of papers on the subject of reflection and transmission 
of a plane electromagnetic wave by a moving medium, starting with the pioneering 
papers by Tai and Yeh, the refraction formulas for different 
setups were derived by using the notions from the classical electromagnetic theory and the 
Lorentz transformation \cite{tai1,tai2,yeh1,yeh2,pyati,shiozawa,huang}.
Although these publications also include the cases 
discussed in our Sections III and IV, the resulting refraction formulas 
do not match the refraction formulas in Eq. (\ref{eq:3.12}) and (\ref{eq:4.18}) in 
our paper \cite{yehcomment}. 
This mismatch is due to the fact that Tai, Yeh and their successors considered the 
angles of incidence, reflection and refraction of the light wave as the angles between 
the corresponding propagation vectors and the interface normal. However, 
we have shown that an optical isotropic medium in uniform rectilinear motion 
becomes an optical anisotropic medium. Therefore, the propagation vector in the 
moving medium will, in general, not coincide with the light ray in the moving 
medium. In conclusion, the refraction formulas by Tai, Yeh and their followers do not 
refer to refraction of a light ray.

The approach in this paper can be extended to include more sophisticated 
situations, specifically, by taking into account the dispersion of the material, 
or considering various three-dimensional cases when the plane of incidence is not 
necessary normal to the interface and not parallel to the velocity of the medium. 
We also suggest a repetition of the derivations in Sec. III and IV when the refraction 
occurs at an interface between two uniformly moving non-vacuum optical materials
having different refractive indices in their rest frames of reference.

\begin{acknowledgments}

The authors are indebted to I. Carusotto for constructive comments that substantially 
improved the discussions in Sections IV and V. We also thank J. Hamel, N. Novkovski, 
K. Tren\v{c}evski and V. Urumov for a critical reading of the manuscript, and 
V. Jovanov for helping us with the references. 

\end{acknowledgments}

\appendix
\section{A derivation of Eq. (\ref{eq:4.18}) by using the relativistic velocity transformation
formulas}

The refraction formula in Eq. (\ref{eq:4.18}) can also be obtained by comparing the path
of the photon in the $S'$- frame where the medium is stationary, to the corresponding path 
of the photon in the $S$-frame where the medium is in uniform rectilinear motion.
We first describe the reflection with respect to $S'$-frame.
Let the vacuum-material interface coincide with $x'$-axis, with $y'>0$ being the region of
the material half-space, and $y'<0$ the region of the vacuum half-space. 
Let the photon be incident on the interface from the vacuum half-space in a direction
making a positive angle $\alpha'$ with the interface normal.  
The photon is refracted in the material half-space at an angle $\beta'$ with respect to the 
interface normal, obeying the Snell's law of refraction:
\begin{equation}
\sin\alpha'=n\sin\beta',
\label{eq:A1.1}
\end{equation}
where $n$ is the refractive index of the medium at rest. Taking that the incident photon 
moves at a speed of light in vacuum $c$, and the refracted photon at a speed $c/n$, the 
velocity components of the photon in the $S'$-frame are:
\begin{eqnarray}
v'_{x_1}&=&c\sin\alpha', \label{eq:A1.2}\\
v'_{y_1}&=&c\cos\alpha', \label{eq:A1.3}\\
v'_{x_2}&=&(c/n)\sin\beta', \label{eq:A1.4}\\
v'_{y_2}&=&(c/n)\cos\beta', \label{eq:A1.5}
\end{eqnarray}
where index ``1'' corresponds to the components of the incident photon, and index ``2'' to the 
components of the refracted photon (for simplicity, we use $v'_x$ and $v'_y$ instead of
$v'_{x'}$ and $v'_{y'}$). 
With respect to $S$-frame where the material half-space is moving at a constant speed 
$u$ in the positive $y$-direction, the photon is incident on the moving interface at an 
angle $\alpha$ at the same constant speed $c$ as in the $S'$-frame, but its speed after 
being refracted from the interface is $v_{\beta}$ which is a function of the angle of refraction $\beta$.
Thus, the corresponding velocity components of the photon in $S$-frame are:
\begin{eqnarray}
v_{x_1}&=&c\sin\alpha, \label{eq:A1.6}\\
v_{y_1}&=&c\cos\alpha, \label{eq:A1.7}\\
v_{x_2}&=&v_{\beta}\sin\beta, \label{eq:A1.8}\\
v_{y_2}&=&v_{\beta}\cos\beta. \label{eq:A1.9}
\end{eqnarray}
Dividing Eq. (\ref{eq:A1.8}) with Eq. (\ref{eq:A1.9}), and using the relativistic transformation 
formulas for the velocity components between $S$ and $S'$:
\begin{eqnarray}
v_x&=&{v'_x(1-u^2/c^2)^{1/2}\over 1+v'_yu/c^2}, \label{eq:A1.10}\\
v_y&=&{v'_y+u \over 1+v'_yu/c^2}, \label{eq:A1.11}
\end{eqnarray}
we obtain:
\begin{equation}
\tan\beta={v'_{x_2}(1-u^2/c^2)^{1/2}\over v'_{y_2}+u}.
\label{eq:A1.12}
\end{equation}
Combining Eqs. (\ref{eq:A1.4}) and (\ref{eq:A1.5}) with Eq. (\ref{eq:A1.1}), we obtain:
\begin{eqnarray}
v'_{x_2}&=&(c/n^2)\sin\alpha', \label{eq:A1.14}\\
v'_{y_2}&=&(c/n)(1-\sin^2\alpha'/n^2)^{1/2}, \label{eq:A1.13}
\end {eqnarray}
which we substitute into Eq. (\ref{eq:A1.12}) to get:
\begin{equation}
\tan\beta={(c/n^2)\sin\alpha'(1-u^2/c^2)^{1/2}\over (c/n)(1-\sin^2\alpha'/n^2)^{1/2}+u}.
\label{eq:A1.15}
\end{equation}
Using the velocity transformation formula:
\begin{equation}
v'_x={v_x(1-u^2/c^2)^{1/2}\over 1-v_yu/c^2}
\label{eq:A.16}
\end{equation}
and Eqs. (\ref{eq:A1.2}), (\ref{eq:A1.6}) and (\ref{eq:A1.7}), we get:
\begin{equation}
\sin\alpha'={\sin\alpha(1-u^2/c^2)^{1/2}\over 1-(u/c)\cos\alpha}.
\label{eq:A1.17}
\end{equation}
We substitute Eq. (\ref{eq:A1.17}) into Eq. (\ref{eq:A1.15}) and simplify the result 
to obtain the law of refraction of the photon in Eq. (\ref{eq:4.18}).

\section{Conditions for the appearance of the backward refraction}

Because a backward refraction occurs only for $u/c<0$, we rewrite Eq. (\ref{eq:4.18}) by
putting $-\kappa$ instead of $u/c$ ($\kappa=\vert u/c\vert$):
\begin{eqnarray}
\tan\beta=(1-\kappa^2)\sin\alpha\times \biggl\{-\kappa n^2\left(1+\kappa\cos\alpha\right) \nonumber \\
+\Bigl[n^2\left[1+\kappa\cos\alpha\right]^2-(1-\kappa^2)\sin^2\alpha\Bigr]^{1/2}\biggr\} ^{-1}. 
\label{eq:A2.1}
\end{eqnarray}
The photon will be refracted at $\beta=90^\circ$ when $\alpha=\alpha_c$, which is the case when
the denominator in the right-hand side of Eq. (\ref{eq:A2.1}) equals zero:
\begin{eqnarray}
\Bigl[n^2\left[1+\kappa\cos\alpha_c\right]^2-(1-\kappa^2)(1-\cos^2\alpha_c)\Bigr]^{1/2}= \nonumber\\
=\kappa n^2\left(1+\kappa\cos\alpha_c\right).
\label{eq:A2.2}
\end{eqnarray}
The last equation is an irrational equation in $\cos\alpha_c$, and its solution 
is equivalent to the solution of the system:
\begin{equation}
\left\{ 
\begin{array}{rcl}
\Bigl[1-\kappa^2+\kappa^2n^2\left(1-\kappa^2n^2\right)\Bigr]\cos^2\alpha_c&&\\
+2\kappa n^2\left(1-\kappa^2n^2\right)\cos\alpha_c&& \\
+1-\kappa^2-n^2\left(1-\kappa^2n^2\right)&=&0 \\
\kappa n^2\left(1+\kappa\cos\alpha_c\right)&\geq&0
\end{array}
\right.
\label{eq:A2.3}
\end{equation}
which simplifies to:
\begin{equation}
\left\{
\begin{array}{rcl}
(\cos\alpha_c)_{1,2}&=&\Bigl\{-\kappa n^2(1-\kappa^2n^2)\pm(1-\kappa^2) \\
& &\times\bigl[1-n^2(1-\kappa^2n^2)\bigr]^{1/2}\Bigr\} \\
& &\times\left[1-\kappa^2+\kappa^2n^2(1-\kappa^2n^2)\right]^{-1} \\
\cos\alpha_c&\geq& -1/\kappa
\end{array}
\right.
\label{eq:A2.4}
\end{equation}
The negative values for $\cos\alpha_c$ imply $\alpha_c>90^\circ$. Since these
solutions are physically unacceptable, we reject the solution with the ``--'' 
sign in the first equation of the system (\ref{eq:A2.4}), because this solution 
corresponds to $\cos\alpha_c<0$. 
What is left to be further analyzed is the expression:
\begin{eqnarray}
\cos\alpha_c&=&\Bigl\{-\kappa n^2(1-\kappa^2n^2)+(1-\kappa^2) \nonumber \\
& &\times\bigl[1-n^2(1-\kappa^2n^2)\bigr]^{1/2}\Bigr\} \nonumber \\
& &\times\left[1-\kappa^2+\kappa^2n^2(1-\kappa^2n^2)\right]^{-1}.
\label{eq:A2.5}
\end{eqnarray}
The right-hand side in Eq. (\ref{eq:A2.5}) is a real number if the expression under
the square-root is non-negative, i.e. $1-n^2(1-\kappa^2n^2)\geq 0$. Hence
$\kappa\geq(n^2-1)^{1/2}/n^2=\kappa_{\text{min}}$. Specifically, when $\kappa=\kappa_{\text{min}}$,
Eq. (\ref{eq:A2.5}) reduces to $\cos\alpha_c=-(n^2-1)^{1/2}/n^2$. Since
$0^\circ\leq\alpha_c\leq 90^\circ$, the right-hand side in Eq. (\ref{eq:A2.5}) should
satisfy the inequality:
\begin{eqnarray}
0&\leq&\Bigl\{-\kappa n^2(1-\kappa^2n^2)+(1-\kappa^2) \nonumber \\
& &\times\bigl[1-n^2(1-\kappa^2n^2)\bigr]^{1/2}\Bigr\} \nonumber \\
& &\times\left[1-\kappa^2+\kappa^2n^2(1-\kappa^2n^2)\right]^{-1}\leq 1.
\label{eq:A2.6}
\end{eqnarray}
For a fixed $n$, one can determine the interval of values for $\kappa$, 
$\kappa\in[\kappa_1, \kappa_2]$, such that for each $\kappa$ within this
interval, Eq. (\ref{eq:A2.5}) has a physically acceptable solution for $\alpha_c$.
This is the critical incident angle at which the backward refraction 
actually appears, and for $\alpha<\alpha_c$ the refraction is regular (forward), 
while for $\alpha>\alpha_c$ the refraction is backward. The ends of the
interval $[\kappa_1, \kappa_2]$ correspond to the critical incident angles 
$\alpha_c=90^\circ$ and $\alpha_c=0^\circ$, respectively, and can be obtained 
from the following equations:
\begin{eqnarray}
\left(1-\kappa_1^2\right)\left[1-n^2(1-\kappa_1^2 n^2)\right]^{1/2}=\kappa_1^2 n^2\left(1-\kappa_1^2 n^2\right), \nonumber\\
\label{eq:A2.7} \\
\left[1-n^2(1-\kappa_2^2 n^2)\right]^{1/2}={1-\kappa_2\left[1-n^2\left(1-\kappa_2^2 n^2\right)\right]\over 1-\kappa_2}. \nonumber \\
\label{eq:A2.8}
\end{eqnarray}
The last two equations are irrational equations in $\kappa_{1,2}$. To avoid cumbersome 
calculations, we substitute $x'=n^2(1-\kappa_1^2n^2)$ 
in Eq. (\ref{eq:A2.7}) and $x''=1-n^2(1-\kappa_2^2n^2)$ in Eq. (\ref{eq:A2.8}) to obtain: 
\begin{eqnarray}
\left(1-x'\right)^{1/2}&=&{\kappa_1x'\over 1-\kappa_1^2}, \label{eq:A2.9} \\
x''^{1/2}&=&{1-\kappa_2x''\over 1-\kappa_2}. \label{eq:A2.10}
\end{eqnarray}
Equation (\ref{eq:A2.9}) is equivalent to the following system:
\begin{equation}
\left\{ 
\begin{array}{rcl}
\kappa_1^2x'^2+\left(1-\kappa_1^2\right)x'-1+\kappa_1^2&=&0 \\
x'&\geq&0
\end{array}
\right.
\label{eq:A2.11}
\end{equation}
while Eq. (\ref{eq:A2.10}) corresponds to:
\begin{equation}
\left\{
\begin{array}{rcl}
\kappa_2^2x''^2-\left(1+\kappa_2^2\right)x''+1&=&0 \\
x''&\leq&1/\kappa_2
\end{array}
\right.
\label{eq:A2.12}
\end{equation}
The quadratic equation in Eq. (\ref{eq:A2.11}) has two solutions in $x'$:
$x'_1=1-\kappa_1^2$ and $x'_2=-(1-\kappa_1^2)/2\kappa_1^2$. The 
solution $x'_2$ is not consistent with the inequality in Eq. (\ref{eq:A2.11}),
since $x'_2<0$ (we have taken into account that $0<\kappa<1$). Thus, the only 
solution of the system (\ref{eq:A2.11}) is $x'_1=1-\kappa_1^2$, leading to the
equation $n^2(1-\kappa_1^2n^2)=1-\kappa_1^2$, which gives:
\begin{equation}
\kappa_1=(1+n^2)^{-1/2}.
\label{eq:A2.13}
\end{equation}
Similarly, by solving the quadratic equation in Eq. (\ref{eq:A2.12}) one obtains
two solutions in $x''$: $x''_1=1$ and $x''_2=1/\kappa_2^2$. Since the solution $x''_2$
does not satisfy the inequality in Eq. (\ref{eq:A2.12}), we reject it and consider
only the solution $x''_1=1$, which leads to $1-n^2(1-\kappa_2^2n^2)=1$, 
and hence:
\begin{equation}
\kappa_2=1/n.
\label{eq:A2.14}
\end{equation}
Equations (\ref{eq:A2.13}) and (\ref{eq:A2.14}) determine the length of the interval 
$[\kappa_1, \kappa_2]$ and its dependence on $n$. We can now summarize the results. 
For $\kappa<(1+n^2)^{-1/2}$, the refraction is regular (forward) 
for any incident angle $\alpha$. When $(1+n^2)^{-1/2}\leq\kappa\leq 1/n$, 
there exists a critical incident angle $\alpha_c$, such that for $\alpha$ in 
the interval $0^\circ<\alpha<\alpha_c$ the refraction is regular (forward), 
while for $\alpha_c<\alpha<90^\circ$ the refraction is backward. The angle 
$\alpha_c$ can be calculated from Eq. (\ref{eq:A2.5}). It depends on $\kappa$, 
and tends to zero as $\kappa$ approaches $1/n$. When $\kappa=1/n$, 
$\alpha_c=0^\circ$, and the right-hand side of Eq. (\ref{eq:A2.1}) 
becomes an undetermined expression of type $0/0$, which can be resolved by 
using L'Hospital's rule. When $\kappa>1/n$, the denominator in Eq. (\ref{eq:A2.1}) 
is negative for any $\alpha$. On the other hand, the numerator in Eq. (\ref{eq:A2.1}) 
is always non-negative. Therefore, for $\kappa>1/n$, $\beta>90^\circ$
for each $\alpha$. Hence, when the medium is moving in the negative direction of the 
$y$-axis, there exists a critical speed for the medium $\vert u_c\vert=\kappa_2 c=c/n$,
corresponding to the treshold for the ``superluminal'' speeds. 
Beyond this critical speed, the photon will always be backwardly refracted.


\begin{thebibliography}{99}
\bibitem{hecht2}
We note that the original statement of Fermat's principle is not 
quite correct, since there are some cases in which the transit time 
is a saddle point. See, e.g., Ref. \cite{hecht}.
\bibitem{sastry2}
Although the speed of each photon on the wavefront of the pulse
will be frame-invariant, the frequency, the intensity and the angular
distribution of the photons will depend on the reference frame
from where the observation is made. See, e.g., Ref. \cite{sastry}.
\bibitem{negative}
The values for the angles in a given half-space are taken positive/negative
if they are in the clockwise/counterclockwise direction from the interface
normal at that particular half-space.
\bibitem{yehcomment}
For example, in the case when the material half-space is moving parallel 
to the interface, Yeh's refraction formula predicts that when the incident 
angle is zero, the refracted angle also will be zero for any value for the speed 
of the medium (see Ref. \cite{yeh1} above). This seems contrary to the observed 
transverse Fresnel-Fizeau light drag, which suggests non-zero angles of refraction 
even when the incident angle is zero. Actually, from an electrodynamics viewpoint,
Yeh considered the phase velocity of light. In this sense, because of momentum
conservation along the interface, an incident beam at normal incidence must refract
into a refracted beam whose phase velocity is also normal to the inerface.
On the other hand, the formulas in Eq. (\ref{eq:3.12}) and (\ref{eq:4.18}) above
are describing the refraction of a light ray. The light ray corresponds to the group 
velocity, and correctly describes the path followed by light if the experiment is 
performed with a pencil of laser light. See, e.g., Ref. \cite{bolotovskii} and
\cite{dispersive}.
\end{thebibliography}
\end{document}